\documentclass{PoS}
\usepackage{graphicx}
\usepackage{epstopdf}
\usepackage{amsmath}

\title{Evolution of multiplicity fluctuations in heavy ion collisions}

\ShortTitle{Evolution of multiplicity fluctuations in heavy ion collisions}

\author{\speaker{Radka Sochorov\'a}\\
	\v{C}esk\'e vysok\'e u\v{c}en\'i technick\'e v Praze, FJFI, B\v{r}ehov\'a 7, 115 19 Praha 1, Czech Republic\\
	E-mail: \email{sochorad@fjfi.cvut.cz}}

\author{Boris Tom\'a\v{s}ik\\
	Univerzita Mateja Bela, Tajovsk\'eho 40, 97401 Bansk\'a Bystrica, Slovakia\\
	\v{C}esk\'e vysok\'e u\v{c}en\'i technick\'e v Praze, FJFI, B\v{r}ehov\'a 7, 115 19 Praha 1, Czech Republic\\
	E-mail: \email{boris.tomasik@fjfi.cvut.cz}}

\abstract{The evolution of multiplicity distribution of a species which undergoes chemical reactions can be described with the help of a master equation.  We study the master equation for a fixed temperature, because we want to know how fast different moments of the multiplicity distribution approach their equilibrium value. We particularly look at the 3rd and 4th factorial moments and their equilibrium values from which central moments, cumulants and their ratios can be calculated. Then we study the situation in which the temperature of the system decreases.   We find out that in the non-equilibrium state, higher factorial moments differ more from their equilibrium values than the lower moments and that the behaviour of the combination of the central moments depends on the combination we choose. If one chooses to determine the chemical freeze-out temperature from the measured values of higher moments, these effects might jeopardise the correctness of the extracted value. }

\FullConference{Corfu Summer Institute 2018 "School and Workshops on Elementary Particle Physics and Gravity"\\
		(CORFU2018)\\
		31 August - 28 September, 2018\\
		Corfu, Greece}

\begin{document}

\section{Motivation}

The main motivation of this work is that measured moments of the multiplicity distribution for various sorts of particles are used for the determination of the hadronisation parameters of hot QCD matter in ultrarelativistic heavy-ion collisions. We assume that there are still some inelastic scatterings after hadronisation that may drive the multiplicity distribution out of equilibrium.
We demonstrate how the different moments depart away from their equilibrium values. If such moments were measured and interpreted as if they were equilibrated, we would obtain different apparent temperatures from different moments. For the description of the evolution of the multiplicity distribution we use a master equation. Because our aim is to study the fluctuations of multiplicities, we actually 
study an ensemble of fireballs and the time evolution of the multiplicity distribution across the ensemble.

\section{Relaxation of factorial moments}
\label{s:2}

We  consider a binary reversible process $a_1 a_2 \longleftrightarrow b_1 b_2 $ with $a \neq b$. Such a reaction  is relevant 
for the investigation of rare species production. 
We  note that the involved species are not identical to each other and it is important to say that $b$-particles carry conserved charge while 
$a$-particles do not. We will also assume that we have a sufficiently large pool of $a$-particles. The pool basically does not change during this chemical process.

Now we can write the master equation \cite{1} for $P_n (t)$, the probability of finding $n$ pairs $b_1 b_2$. It has the following form 
\begin{equation}\label{eq:1}
\dfrac{dP_n(t)}{d t} = \dfrac{G}{V} \left\langle N_{a_1} \right\rangle  \left\langle N_{a_2} \right\rangle  \left[ P_{n-1}(t) - P_n(t)\right] -  \dfrac{L}{V} \left[ n^2 P_n(t) - \left( n+1\right) ^2 P_{n+1}(t)\right],
\end{equation} 
where $n = 0, 1, 2, 3, ...$ and $V$ is proper volume of the reaction. For a thermal distribution of particle momentum, $G \equiv \langle \sigma_G v \rangle $ is gain term (describes creation) and $ L \equiv \langle \sigma_L v \rangle $ is loss term (describes annihilation). These two terms are averaged cross-sections. 

The probability $P_n$ 
which is described by eq. (\ref{eq:1}) increases when a pair of $b_1b_2$ is produced from the state with $(n-1)$ pairs or a pair is destroyed from 
the state with $(n+1)$ pairs. On the other hand, it decreases with creation or annihilation of a pair from the state with $n$ pairs.

When we want to study thermalisation, then it is useful to cast the equation into dimensionless form 
with the help of the time variable $\tau = t L/V$, where $V/L = \tau_{0}^{c}$ is so-called relaxation time. 
Now, $G, V $ and $L$ are constant so the master equation formulated in this dimensionless time must describe the approach towards equilibrium. 
In terms of the relaxation time, the evolution is universal and same for all reactions. Instead of the constants $G, L,  \left\langle N_{a_1} \right\rangle $ 
and $ \left\langle N_{a_2} \right\rangle$ we use $ \epsilon = G  \left\langle N_{a_1} \right\rangle  \left\langle N_{a_2} \right\rangle / L $. 

The master equation can be converted into a partial differential equation for a generating function \cite{1}
\begin{equation}\label{eq:2}
g(x, \tau) = \sum_{n=0}^{\infty} x^n P_n (\tau),
\end{equation} 
where $x$ is an auxiliary variable.

From the derivative of the generating function we can determine the equilibrium values of the factorial moments.

If we multiply eq. (\ref{eq:1}) by $x^n$ and sum over $n$, we find that \cite{1} 
\begin{equation}
\frac{\partial g(x, \tau)}{\partial \tau}= (1-x)(xg''+g'-\epsilon g),
\end{equation}
where $g' = \partial g/\partial x$. The generating function obeys the normalisation condition
\begin{equation}
g(1, \tau) = \sum_{n=0}^{\infty} P_n (\tau) = 1.
\end{equation}
The equilibrium solution, $g_{eq} (x)$, must not depend on time, thus it obeys the following equation
\begin{equation}
x g_{eq} ^{''} + g_{eq} ^{'} - \epsilon g_{eq} = 0.
\end{equation} 
The solution that is regular at $x = 0$ is then given by
\begin{equation}
g_{eq} (x) = \frac{I_0 (2 \sqrt{\epsilon x})}{I_0 (2 \sqrt{\epsilon})}.
\end{equation}
 
The average number of $b_1 b_2$ pairs per event in equilibrium is given by 
\begin{equation}
\left\langle N \right\rangle _{eq} = g_{eq}^{'}(1) = \sqrt{\epsilon} \dfrac{I_1(2 \sqrt{\epsilon})}{I_0 (2 \sqrt{\epsilon})}.
\end{equation}  
Here, $I_0(x)$ and $I_1(x)$ are the Bessel functions. Higher derivatives give us then the 
equilibrium values of the  factorial moments which are defined as (their scaled values)
\begin{eqnarray}\label{eq:3}
F_2 & = & \frac{\langle N (N-1) \rangle}{\langle N \rangle^2}\\
F_3 & = & \frac{\left \langle N (N-1) (N-2)\right \rangle}{\langle N \rangle^3}\\
F_4 & = & \frac{\left \langle N (N-1) (N-2) (N-3)\right \rangle}{\langle N \rangle^4} \,   .
\end{eqnarray}

Equilibrium values of the 2nd \cite{1,2}, 3rd and 4th \cite{Sochorova:2018ojd} 
factorial moments (not scaled) are then 
\begin{eqnarray}\label{eq:4}
F_{2, eq}  
& = & - \frac{1}{2} \sqrt{\varepsilon}
\frac{I_1(2\sqrt{\varepsilon})}{I_0(2\sqrt{\varepsilon})}  + \frac{1}{2}\varepsilon 
\frac{I_2(2\sqrt{\varepsilon}) + I_0(2\sqrt{\varepsilon})}{I_1(2\sqrt{\varepsilon})} \\
F_{3, eq} 
& = & 
\frac{3}{4} \sqrt{\varepsilon}\frac{I_1(2\sqrt{\varepsilon})}{I_0(2\sqrt{\varepsilon})} \nonumber - \frac{3}{4} \varepsilon\left ( 1 +  \frac{ I_2(2\sqrt{\varepsilon})}{I_0(2\sqrt{\varepsilon})}\right ) \\
&&+ \frac{1}{4} \varepsilon^{3/2} 
\frac{I_3(2\sqrt{\varepsilon}) + 3I_1(2\sqrt{\varepsilon})}{I_0(2\sqrt{\varepsilon})}\\
F_{4, eq} 
& = & -\frac{15}{8} \sqrt{\varepsilon} \frac{I_1(2\sqrt{\varepsilon})}{I_0(2\sqrt{\varepsilon})} \nonumber + \frac{15}{8} \varepsilon \left (\frac{I_2(2\sqrt{\varepsilon})}{I_0(2\sqrt{\varepsilon})} +1\right) \\
&&-\frac{3}{4} \varepsilon^{3/2} 
\frac{3I_1(2\sqrt{\varepsilon}) +I_3(2\sqrt{\varepsilon})}{I_0(2\sqrt{\varepsilon})} 
+ \frac{1}{8} \varepsilon^2 \left ( 3 + 
\frac{4I_2(2\sqrt{\varepsilon}) + I_4(2\sqrt{\varepsilon})}{I_0(2\sqrt{\varepsilon})}\right ) \, .
\end{eqnarray}

Now we study the relaxation of the multiplicity distribution  with the help of the master equation. 
For numerical calculations binomial initial conditions are used
\begin{eqnarray}\label{eq:5}
P_0(\tau = 0) & = & 1-N_0\\
P_1(\tau = 0) & = & N_0\\
P_n(\tau = 0) & = & 0\, \qquad \mbox{for}\phantom{m}n>1\, ,
\end{eqnarray}
where $N_0 = 0.005$ $(N_0 = \left\langle N \right\rangle (\tau = 0))$.

\begin{figure}[t]
	\centerline{\includegraphics[width=0.7\textwidth]{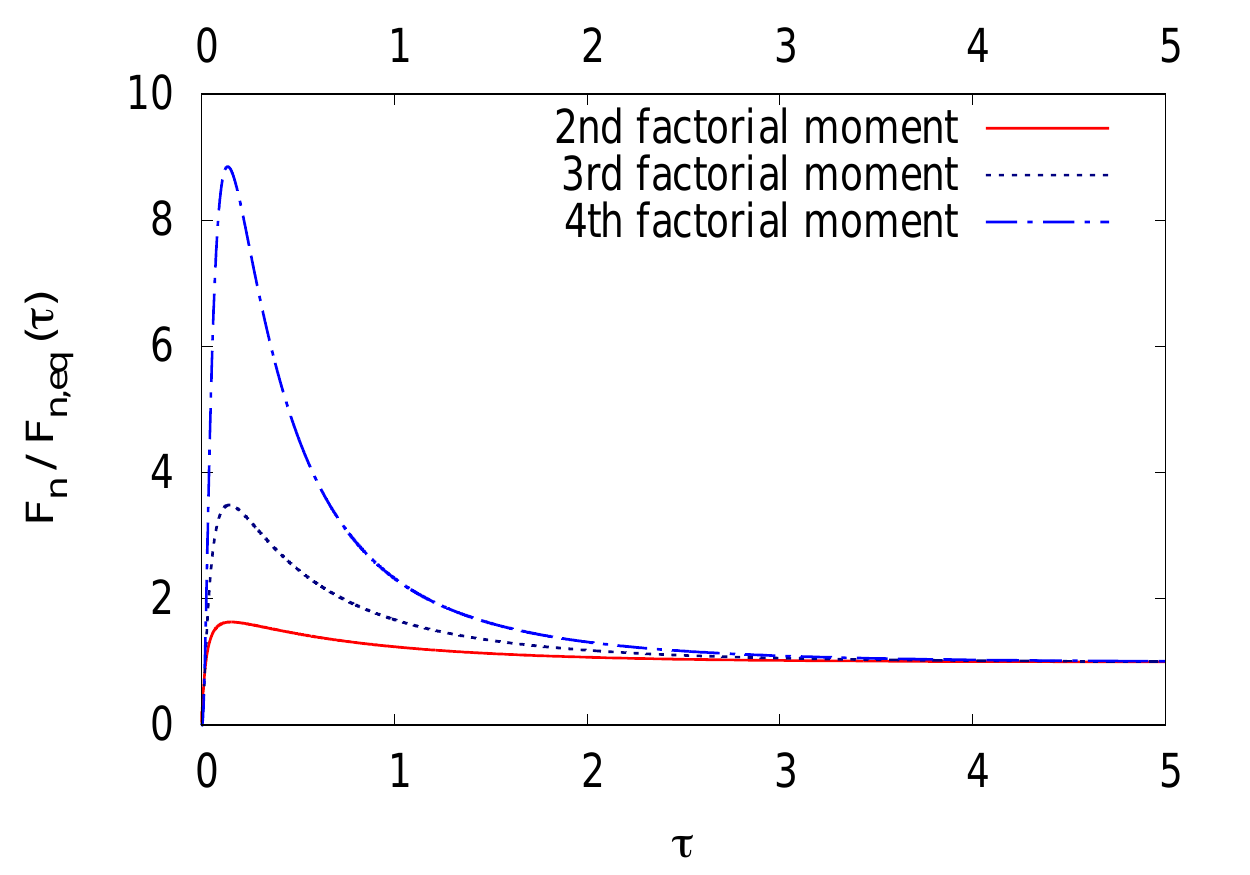}}
	\caption{Time evolution of scaled factorial moments divided by their equilibrium values for
		constant temperature and $\epsilon = 0.1$.}
	\label{fig:1}
\end{figure}

The evolution of the 2nd, 3rd and 4th scaled factorial moments divided by their equilibrium values is shown in Figure \ref{fig:1}.
The value of the parameter $\epsilon$ has been set to 0.1. Note that we obtained qualitatively similar results also for other values of  $\epsilon$. 
We can see that all moments relax within the same dimensionless time $\tau$ and that higher moments differ more from their equilibrium values than lower moments.

	
\section{Higher moments  in a cooling fireball}
\label{s:3}	

Master equation defined in dimensionless time $\tau$ can be used only for constant temperature. 
However, we want to study a more realistic case in which the fireball will cool down. 
A change of temperature implies a variation of the relaxation time.   
Hence, we have to use the original master equation which is defined by eq. (\ref{eq:1}) 
and evaluate the creation and annihilation terms for each temperature. In order to place ourselves into an interesting regime, we have choosen the reaction system $\pi^+ n \longleftrightarrow K^+ \Lambda$. At present we shall use a parametrisation of the cross-section \cite{3}
\begin{equation}\label{eq:10}
\sigma_{\pi N}^{\Lambda K} = \left \{ 
\begin{array}{lc}
0\, \mbox{fm}^2 & \sqrt{s} < \sqrt{s_0}\\
\frac{0.054 (\sqrt{s} - \sqrt{s_0})}{0.091}\, \mbox{fm}^2 &  \sqrt{s_0}\le\sqrt{s}<\sqrt{s_0}+0.09\,\mbox{GeV}\\
\frac{0.0045}{\sqrt{s} - \sqrt{s_0}} \, \mbox{fm}^2 & \sqrt{s} \ge \sqrt{s_0}+0.09\,\mbox{GeV}
\end{array}
\right .
\end{equation}
where $\sqrt{s_0}$ is the threshold energy of the reaction and the energies are given in $\mathrm{GeV}$. 

The evolution starts at $T=165$~MeV. At this temperature, where the hadronisation happens, the system is generated in chemical 
equilibrium. We further calculate how the multiplicity distribution changes. 

We  use a simple toy model in which the temperature and volume behave like in 1D longitudinally boost-invariant expansion (Bjorken scenario).

The temperature drops  according to
\begin{equation}\label{eq:6}
T^3 = T_{0}^{3} \dfrac{t_0}{t}
\end{equation}
all the way down to the final temperature $T = 100~ \mathrm{MeV}$.  
Motivated by  femtoscopic measurements we set the final time to $10 ~\mathrm{fm}/c$. This leads to $t_0 = 2.2 ~ \mathrm{fm}/c$. 	

The effective system volume grows linearly 
\begin{equation}\label{eq:7}
V(t) = V_0 \frac{t}{t_0},
\end{equation}	
where the initial volume we set $V_0 = 125 ~\mathrm{fm^3}$. 

If the chemical processes under investigation are much faster than the characteristic time scale of the expansion, then the multiplicity 
distribution will be always adapted to the ambient temperature. If, on the other hand, chemistry is much slower than the expansion,
then the distribution will barely change. Hence, the interesting regime, where non-equilibrium evolution is expected, is when the reaction 
rate and the expansion rate are roughly of the same order. In order to investigate such a regime, we
scale up the cross-section and in the next part of this work we also investigate the influence of density dependence of the masses. 

\begin{figure}[t]
	\centerline{\includegraphics[width=0.7\textwidth]{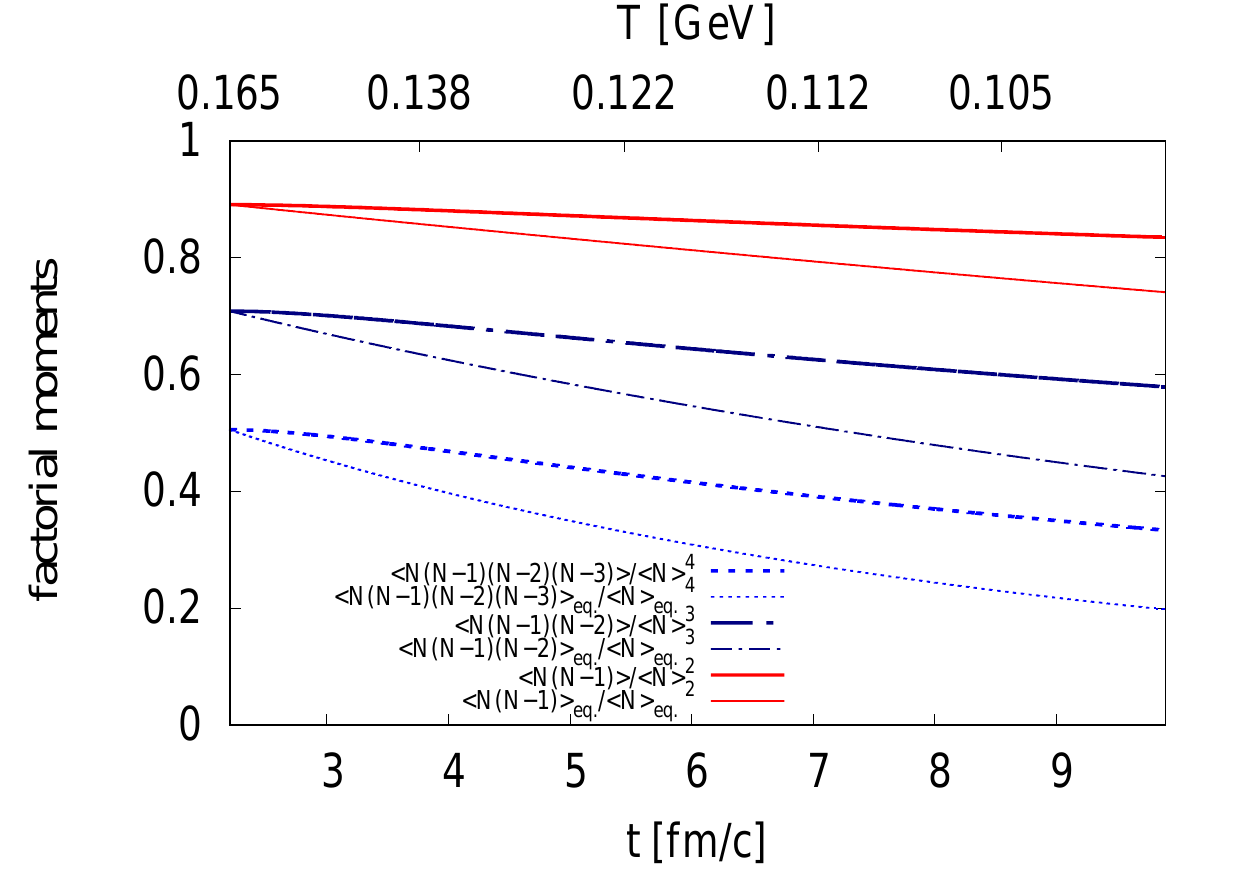}}
	\caption{Evolution of scaled factorial moments for gradual decrease of the temperature from $165~ \mathrm{MeV}$ to $100 ~\mathrm{MeV} $. This graph is plotted for 15 pions and 10 neutrons and for 50 times enlarged cross-section. Thick lines represent the evolution of moments according to the master equation and thin lines represent the equilibrium values calculated for each temperature.}
	\label{fig:2}
\end{figure}

The evolution of scaled factorial moments for a gradual decrease of the temperature is shown in Figure \ref{fig:2}.
We can see that since the temperature is decreasing, the moments change, but the reaction rate is too low to keep them in equilibrium.


\section{The apparent freeze-out temperature}
\label{s:4}

\begin{figure}[t]
	\centerline{\includegraphics[width=0.7\textwidth]{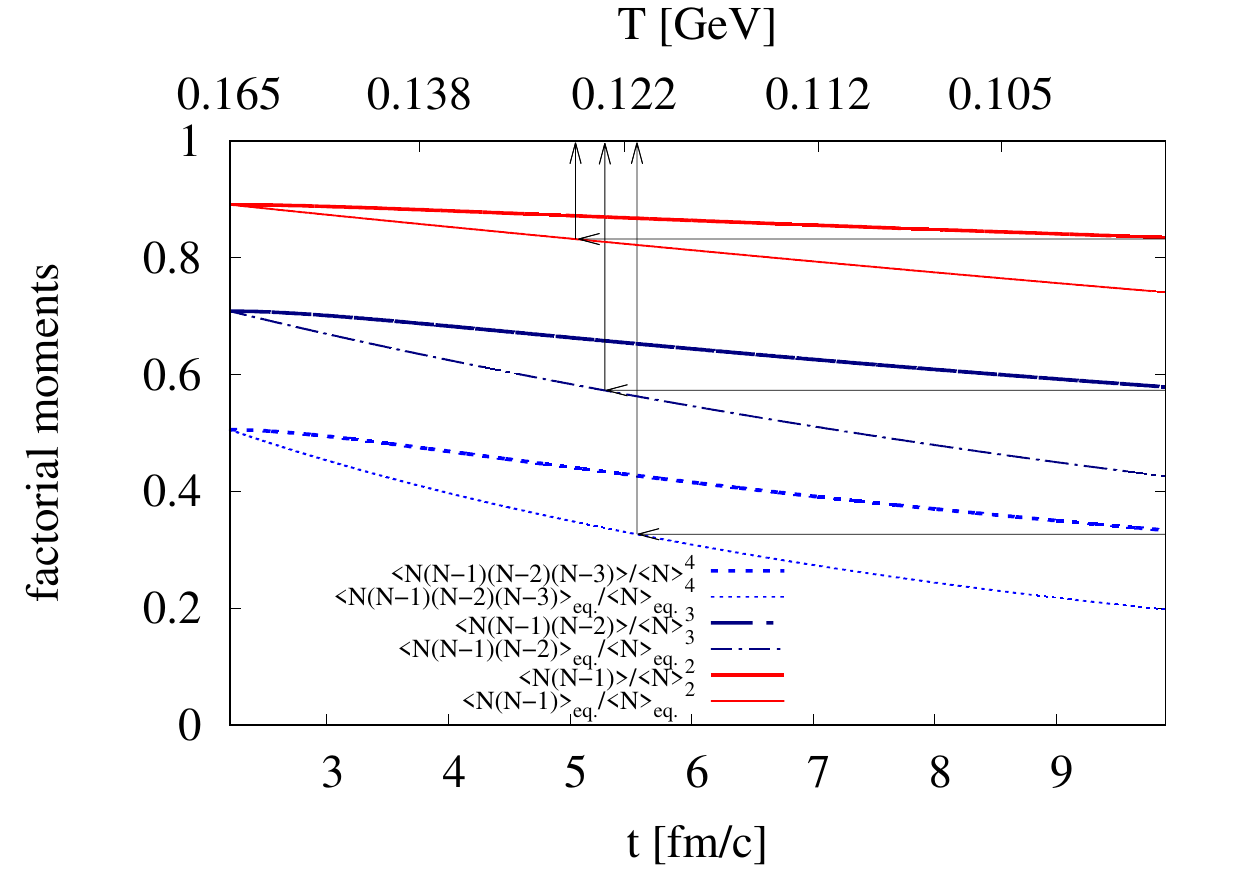}}
	\caption{Evolution of scaled factorial moments for gradual change of temperature from $165~ \mathrm{MeV}$ to $100 ~\mathrm{MeV} $. This graph is plotted for 15 pions and 10 neutrons and for 50 times enlarged cross-section. Thick lines represent the evolution of moments according to the master equation and thin lines represent equilibrium values calculated for each temperature.}
	\label{fig:3}
\end{figure}

We can now demonstrate the potential danger in case of extraction of the freeze-out temperature from the different moments. Suppose 
that we observe the final values of factorial moments that the system eventually achieves in its non-equilibrium evolution, 
as shown in Figure~\ref{fig:3}. 
Suppose further, that we wrongly assume that  
the system is still thermalised. 
This would mean that it has evolved along the thin lines in Fig.~\ref{fig:3}. 
Now we  ask at what temperature would a thermalised system lead to the observed value of the factorial moments. 
The actual observed final value of a thick line is thus projected horizontally on the corresponding thin line (Figure \ref{fig:3}) and the 
apparent temperature is read off from the abscissa.  We can see that such a procedure can lead to different values of the 
apparent temperature if different moments are used. We can also see that higher factorial moments 
seem to indicate lower temperatures than lower moments.

In data analysis, central moments are often used, which are defined as 
\begin{eqnarray}\label{eq:12}
\mu_1 & = & \langle N \rangle = M\\
\mu_2 & = & \langle N^2 \rangle - \langle N\rangle^2 = \sigma^2\\
\mu_3 & = & \langle (N-\langle N\rangle )^3\rangle \\
\mu_4 & = & \langle (N-\langle N\rangle )^4\rangle.
\end{eqnarray}
Often, one uses their combinations, like the  skewness
\begin{equation}\label{eq:13}
S = \frac{\mu_3}{\mu_2^{3/2}} 
\end{equation}
or the  kurtosis
\begin{equation}\label{eq:14}
\kappa = \frac{\mu_4}{\mu_2^2} - 3.
\end{equation}

\begin{figure}[!t]
	\centerline{\includegraphics[width=0.57\textwidth]{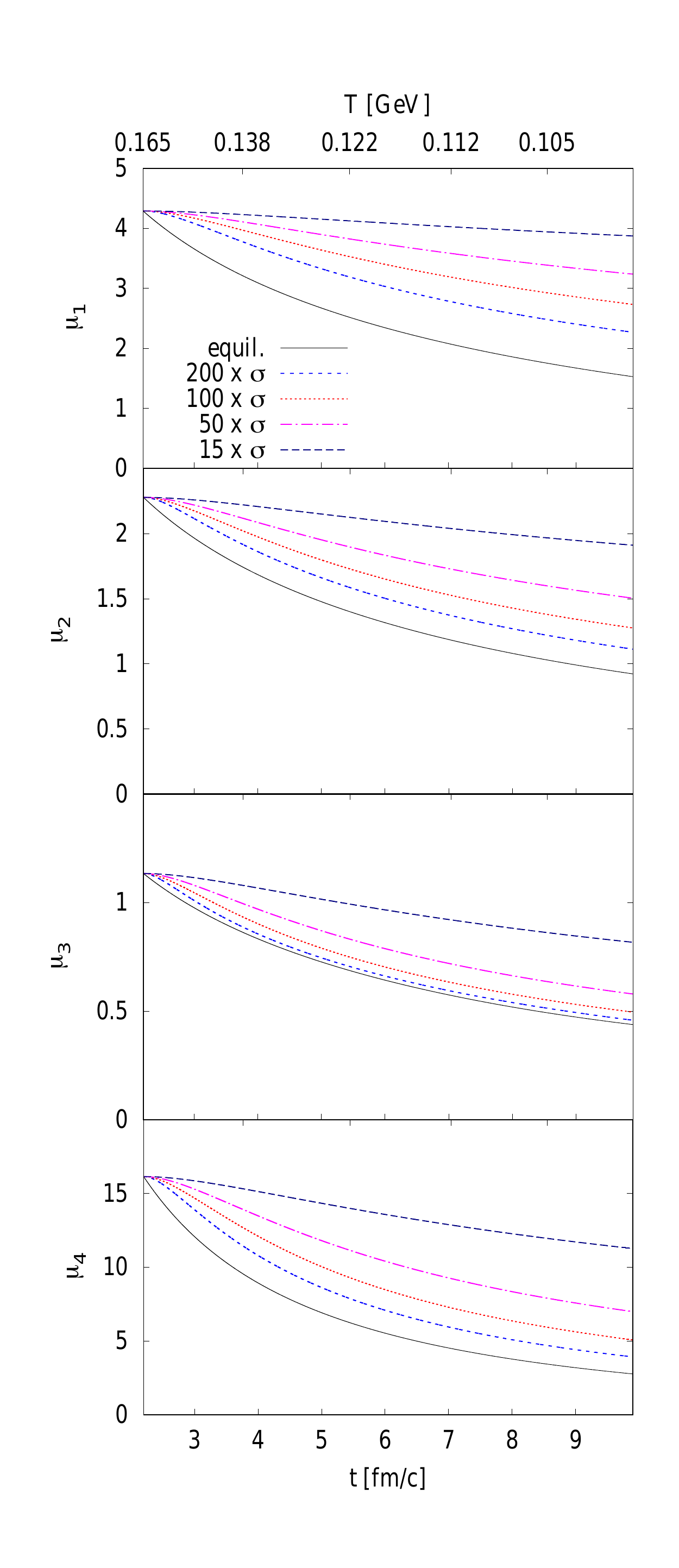}}
	\caption{Evolution of the first four central moments (from top to bottom). Different curves (different colours) on the same panel 
	show results for cross sections scaled by different factors. Solid lines represent the equilibrium values.}
	\label{fig:4}
\end{figure}

We thus investigate the evolution of the  
central moments and 
their combinations when the fireball cools down. 
It turns out that  it is very difficult to extract the exact freeze-out temperature from the non-equilibrium values.

\begin{figure}[t]
	\centerline{\includegraphics[width=0.65\textwidth]{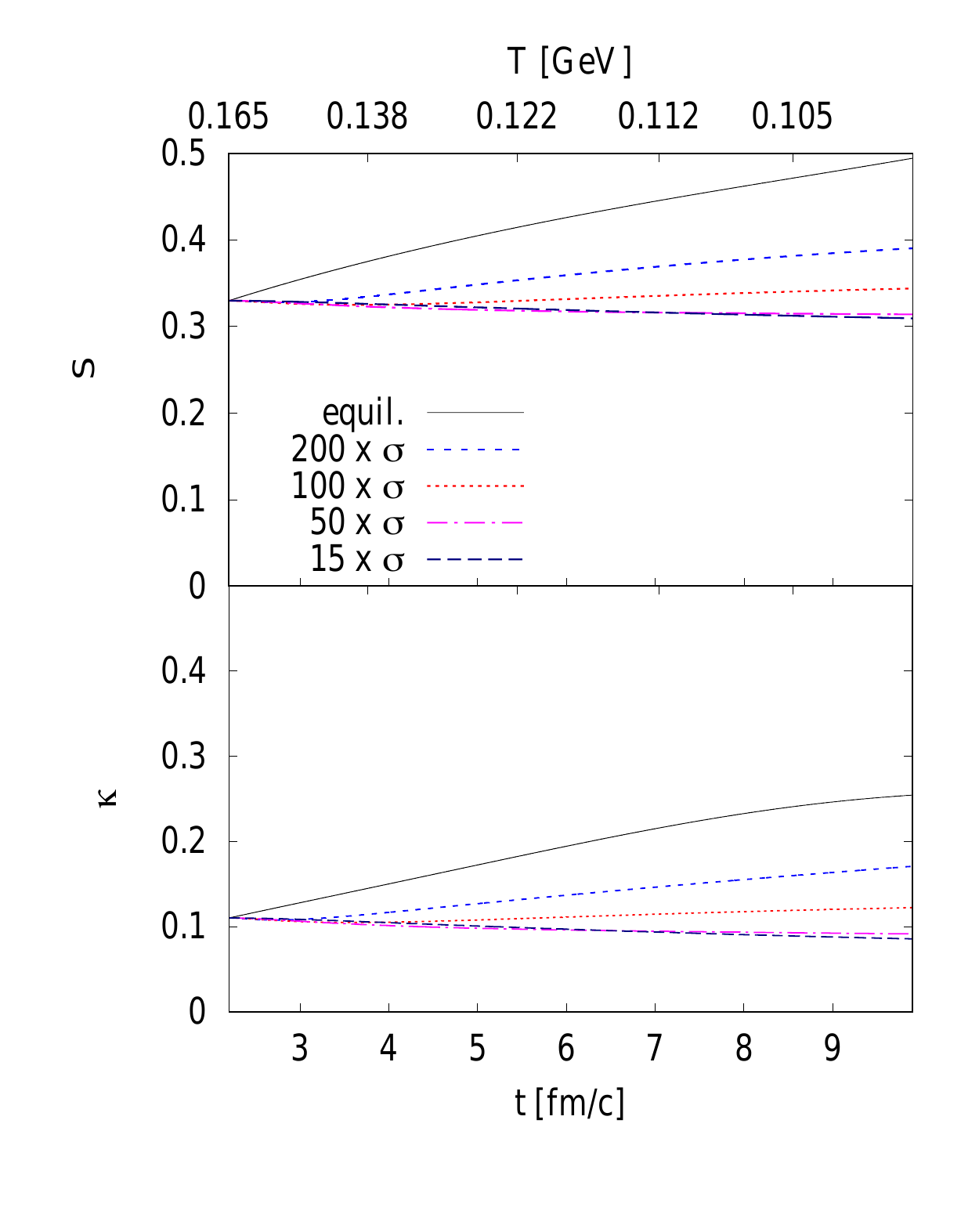}}
	\caption{Evolution of the skewness (upper panel) and the kurtosis (lower panel). Different curves (different colours) on the same panel show results for different cross sections. Solid lines represent the equilibrium values.}
	\label{fig:5}
\end{figure}
We can see in Figure \ref{fig:4} and Figure \ref{fig:5} that while the central moments are decreasing, 
the coefficients of skewness and kurtosis are increasing in the scenario of temperature decrease due to boost-invariant expansion.


\section{Decreasing mass of $\Lambda$ hyperon}
\label{s:5}

In the previous part of this work we assumed that the involved masses and cross-section do not depend on density. 
The gain term of our reaction $\pi^+ + n \longleftrightarrow K^+ + \Lambda$ is small because of the rather higher threshold, which is about $530 ~\mathrm{MeV}$ above the masses of the incoming particles, while the temperature is lower than $165 ~\mathrm{MeV}$. It means that the reaction rate might increase if the threshold is lowered, for example through a decrease of the hyperon mass in baryonic matter. 

Therefore, we now explore the possibility of decreasing the $\Lambda$ hyperon mass.
It means that also the threshold for the reaction is lowered and its rate may grow due to the increase of the available phase space.

We  assume a simple  dependence of the hyperon mass on baryon density $\rho_B$
\begin{equation}
m_{\Lambda} (\rho_B) = \dfrac{\rho_0 - \rho_B}{\rho_0} m_{\Lambda 0} + \dfrac{\rho_B}{\rho_0} m_{p0}.
\end{equation} 
Note that the hyperon mass becomes identical to that of proton $m_{p0}$
at the highest baryon density $\rho_0$ at which our calculation starts. Hyperon mass returns to the vacuum value $m_{\Lambda 0}$ 
if baryon density vanishes. 

Density in our toy model also evolves according to one-dimensional  longitudinally boost-invariant expansion 
\begin{equation}
\rho = \rho_0 \dfrac{t_0}{t},
\end{equation}
where $\rho_0 = 0.08 ~\mathrm{fm^{-3}}$.

For this scenario we present the volume-independent ratios which are often measured. These are, e.g.\ 
\begin{eqnarray}
R_{32} & = & \frac{\mu_3}{\mu_2} = S\sigma\\
R_{42} & = & \frac{\mu_4}{\mu_2} - 3\mu_2 = \kappa\sigma^2.
\end{eqnarray}

\begin{figure}[t!]
	\centering
	\includegraphics[width=1\textwidth]{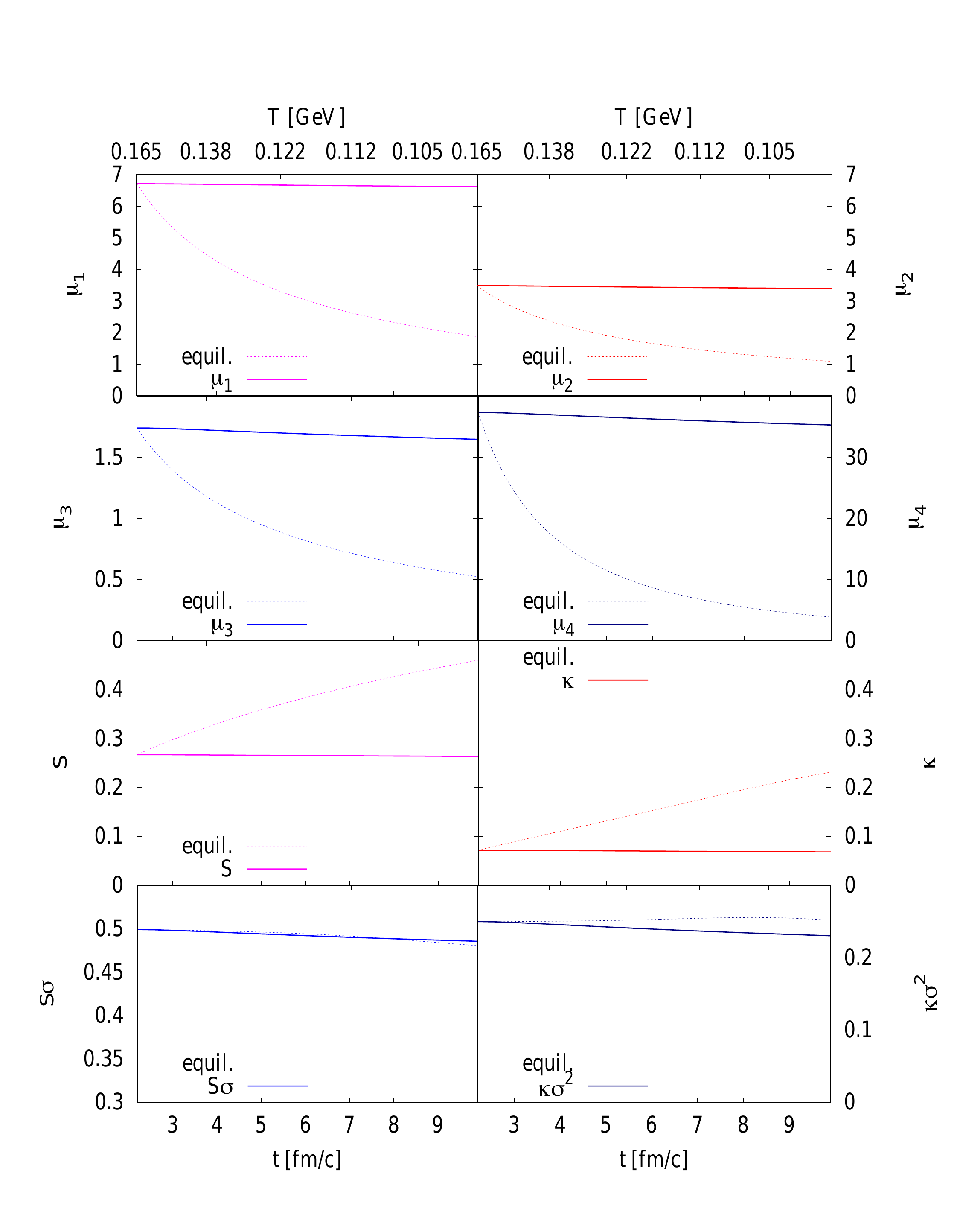}
	\caption{Central moments, skewness, kurtosis and volume-independent ratios $S \sigma$ and $\kappa \sigma^2$ for the scenario with density-dependent mass of $\Lambda$ and the decreasing temperature. Thick solid lines represent numerically calculated evolution and thin dotted lines represent equilibrium values at the given temperature.}
	\label{fig:6}
\end{figure}

Results for the case with density-dependent mass of hyperon are plotted in Figure \ref{fig:6}.

We can see that also for the time evolution of moments and their combinations for density-dependent mass the central moments are decreasing while the coefficients of skewness and kurtosis are increasing. 
Only weak time dependence is 
seen for the volume independent ratios $S \sigma$ and $\kappa \sigma^2$. Thus in real collisions,
where non-equilibrium evolution is likely, it is very  difficult to determine the unique freeze-out temperature from the measured moments.

Nevertheless, in realistic fireballs there are also other channels that can change the numbers of kaons and/or lambdas so we have to expect that moments may change stronger than in our work. 


\section{Conclusion}
\label{s:6}

If chemical equilibrium is broken, higher factorial moments of multiplicity distribution differ more from their equilibrium values than the lower moments. Evolution of chemical reaction off equilibrium may show different temperatures for different orders of the factorial or central moments (or their combinations). We demonstrated this on the reaction $\pi^+ + n \longleftrightarrow K^+ + \Lambda$. The behaviour of the combination of the central moments depends on which combination of moments we choose. 

Hence, one should be very careful when extracting the freeze-out temperature from higher  moments. 


\begin{acknowledgments}
	This work was supported by the grant 17-04505S of the Czech Science Foundation (GA\v{C}R).
	BT also acknowledges the support by VEGA 1/0348/18.
\end{acknowledgments}

\end{document}